# Identified high $p_T$ particle correlation studies in central Au + Au collisions at $\sqrt{s_{NN}} = 200$ GeV

Ying Guo *for the STAR Collaboration*

*Wayne State University*
*Department of Physics & Astronomy 666 West Hancock, Detroit MI 48201*

We present azimuthal two particle correlations of high pt strange baryons and mesons with charged hadrons in Au+Au collisions at $\sqrt{s_{NN}}$ =200 A GeV. We observe the suppression of the back to back correlations similar to the previous measurement of azimuthal correlations of unidentified charged hadrons in the central Au+Au collisions. The dependence of the suppression on the leading particle transverse momentum is presented. We find that Lambda and Anti-Lambda correlations with charged hadrons have different leading hadron $p_T$ dependencies for same side correlations, which could indicate different fragmentation functions for gluon and quark jets or non fragmentation mechanisms for particle production. In addition we show that the back-to-back correlation structure for charged hadron pairs with a high $p_T$ trigger (p$_T$<8 GeV) is consistent with statistical momentum conservation, with no evidence of additional jet-like correlations. *This is a summary of a poster presented at Quark Matter 2004.*

## I. INTRODUCTION:

High $p_T$ two particle correlations have been used to study jet related processes [1]. A two particle analysis does not require the full reconstruction of the jet, which is an advantage in heavy ion collisions where an unambiguous jet reconstruction is difficult due to the large background. In fact in central Au + Au collisions only 0.5% of the charged particles are above $p_T$~2 GeV/c. In two particle high $p_T$ correlations we distinguish between the trigger particle and the associated particle. The trigger particle is a high $p_T$ particle above a certain threshold and it is being associated with another high $p_T$ particle above a defined threshold but below the $p_T$ of the trigger particle. The correlation function shows us the angular distribution of the associated particle with respect to the trigger particle. It can be used to deduce information about the high pt particle interactions with the medium during propagation. This makes identified particle correlation studies interesting since they provide additional information on jet quenching and particle production mechanisms. In particular, recent measurements of the baryon/meson enhancement in the inclusive yields of $K_0^S$ and $\Lambda$ at intermediate $p_T$ ($p_T$~2-6 GeV/c), as shown in M. Lamont's talk at QM04, indicate an interesting behavior of particle identified yields in the high $p_T$ part of the spectra in central Au + Au collisions. Two particle correlations studies will help to disentangle the different mechanisms of particle production in this $p_T$ range.

In this paper, we present two particle correlations of high $p_T$ strange baryons ($\Lambda$, $\overline{\Lambda}$) and mesons ($K_0^S$) with charged hadrons in central Au + Au collision. We observed a similar suppression of the back side correlation as seen in the previous charged hadron measurement, which was attributed to the final state effects [2]. In part A. and B. of this write-up we show the dependence of the suppression on the transverse momentum of the trigger particle in central Au + Au collisions. In part C we presents results of a study of the shape of the back side correlation between unidentified charged hadrons in central Au + Au collisions. The resulting fitting parameters were compared to the estimated contribution from momentum conservation in the system. This study provides us with information about recoil of back side jets inside the media.

## II. ANALYSIS AND RESULTS

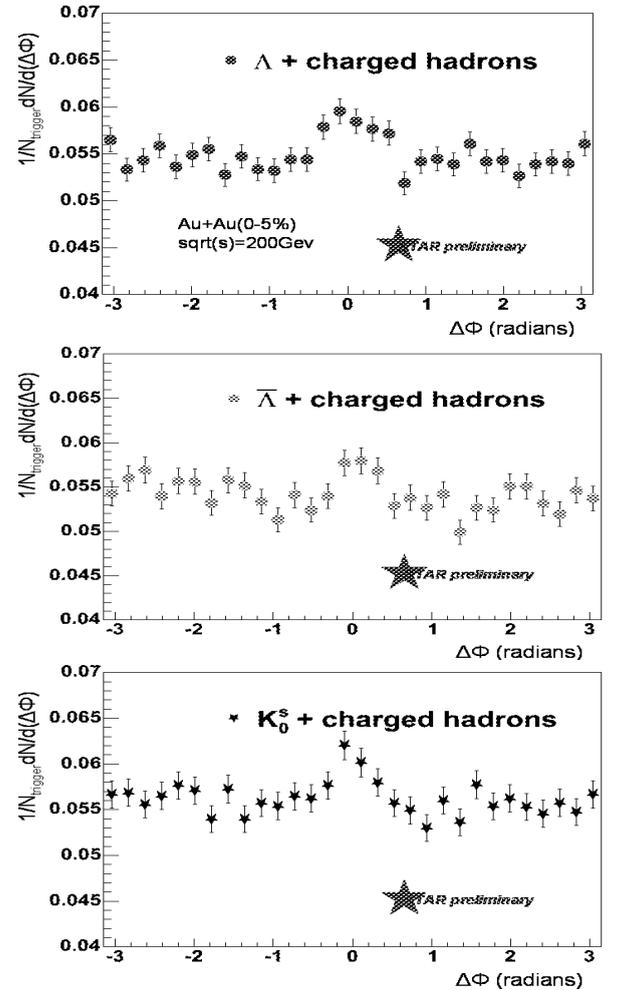

Fig. 1 Correlations of different identified leading particles with charge hadrons. ($p_T$ trigger >2.5 Gev/c, $p_T$ trigger>$p_T$ associated >2.5 GeV/c)

The results shown are based on 800,000 most central events(0-5%) taken with the STAR detector [3]. The $\Lambda, \overline{\Lambda}$ and $K_0^S$ are reconstructed from their decay daughters. This method allows us to reconstruct strange particles to higher $p_T$ without explicit single track particle identification. Detailed information about tracking, efficiency correction, and V0 reconstruction is given in [4,6]. We chose unidentified charged hadrons as the associated particle because the statistics are too limited when requiring both the trigger and the associated particle to be strange particles. In our study the trigger and the associated particles are required to have a transverse momentum greater than 2.5 GeV/c in order to reduce uncorrelated background.

Fig. 1 shows the measured correlations in $\Delta\varphi$ for the different trigger particle species. Generally the high $p_T$ Di-hadron correlation signal is seen still to be present on top of a background, which is generated by uncorrelated processes. The harmonic shape of the background can be attributed to elliptic flow and characterized by the flow parameter $v_2(p_T)$ [7]. To isolate the jet contribution one needs to subtract such effects. One approach is to calculate and subtract the background, the other is to take advantage of the apparent symmetry of the $\Delta\varphi$ distribution and form a difference.

A. Difference function:

We define the difference function (DF) as the difference in number of associated particle in the same and opposite direction of the trigger particle. Let us first introduce two measurement variables $N_{same}$ and $N_{back}$, which are defined as follows. Then we can define the difference function (DF) as the difference of $N_{same}$ and $N_{back}$:

$$DF = N_{same} - N_{back}$$
$$N_{same} = \frac{\sum N_{pairs}(|\Delta\varphi| \leq 0.65)}{N_{trigger}} \quad (1)$$
$$N_{back} = \frac{\sum N_{pairs}(|\Delta\varphi| \geq 2.49)}{N_{trigger}}$$

The difference function gives us the information on the jet quenching effects (difference in number of associated particles between same side jet and back side jet). Due to the forward–backward symmetry of elliptic flow and azimuthal uniformity of the uncorrelated background, the difference functions extract the difference in the correlated yield of the near and back to back hadron pairs. Fig. 2 shows the DF function for two particle correlations for various identified trigger particles as function of trigger $p_T$. The positive values of the DF indicate that back to back correlations are suppressed relative to near side correlations. However the relative contribution of the same side and the back side jet to the difference function can not be unambiguously determined in this method, but can be achieved with the direct background subtraction method which will be discussed in the following section. We also note that the trends of the trigger $p_T$ dependence for the Lambda and Anti-Lambda are slightly different in central collisions, which could be caused by medium effects on different associated particle production mechanisms. However, we need more statistics to confirm the significance of our measurement.

B. Direct background subtraction.

The uncorrelated background of the correlation function can be estimated by using the convolution of the single particle spectra. Because of the full azimuthal coverage of the STAR detector, the calculation is simplified. Equation (2) shows how the random background was calculated. Fig.3 shows the same side and back side integral where we chose the same cone size as in method (A). The suppression effect on the back side correlation is quite obvious and similar for all leading particle species. Therefore the observed difference between species in the difference function is due to the same side correlations.

$$B = \frac{1}{2\pi N_{trigger}} \int_{p_{T,trigger},low}^{p_{T,trigger},high} \frac{dN_{trigger}}{dp_{T,trigger}} dp_{T,trigger} \int_{p_{T,associated},low}^{p_{T,associated},high} \frac{dN_{associated}}{dp_{T,associated}} dp_{T,associated} \quad (2)$$

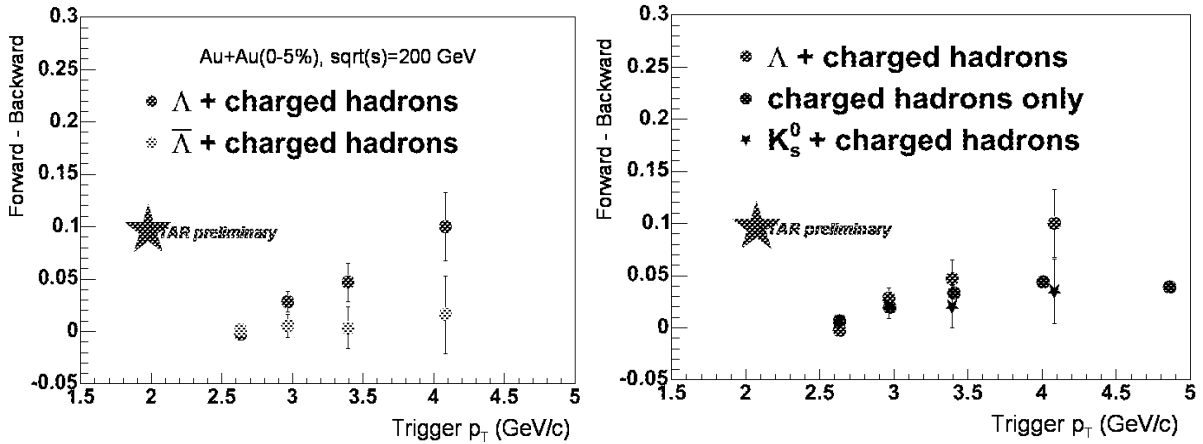

Fig.2 Forward-Backward difference in correlated yield as function of transverse momentum of the trigger particle

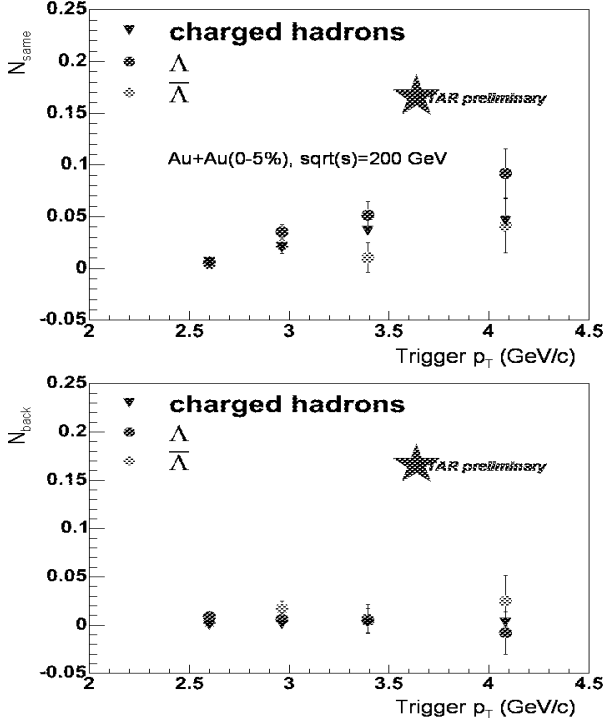

Fig. 3 Same side and back side of jet correlation

## C. Recoil against high $p_T$ charged hadrons trigger:

In central Au+Au collisions, the back side correlation apparently lost most of its energy inside the medium so that we see no punch through effects even for leading trigger particles up to a transverse momentum at 6~8 GeV/c[1,2]. However the momentum has to be conserved. To balance out the transverse momentum carried by the same side correlation, the same amount of momentum on the back side had to be absorbed and redistributed among many particles via recoil processes. In the most extreme case this momentum could get redistributed over the whole system. Previous STAR studies have revealed that the away-side correlation peak of a high $p_T$ hadron recoiling against a charged hadron trigger with 4<$p_T$<6 GeV/c is strongly suppressed in central Au+Au collisions [1,2]. This has been interpreted as the suppression of back-to-back jets, with the jet recoiling against the trigger dissipating its energy in the medium. However, momentum must be conserved. It is an open question whether full dissipation occurs in the medium, with momentum distributed over many particles, or whether jet-like correlations persist for the recoiling object.

Borghini et al [5] provide a means to investigate this question. In the case of statistical (uncorrelated) momentum conservation, the two particle correlation has the form of equation (3):

$$C^{\Sigma P_T}(\mathbf{p}_1,\mathbf{p}_2) = \frac{f_c(\mathbf{p}_1,\mathbf{p}_2)}{f_c(\mathbf{p}_1)f_c(\mathbf{p}_2)} - 1 = -\frac{2\mathbf{p}_{T1}\mathbf{p}_{T2}}{N<P_T^2>} \quad (3)$$

Insofar as the angular shape and $p_T$ dependence of the away-side peak are described by equation (3), there is no evidence for dynamical correlations in the momentum balance.

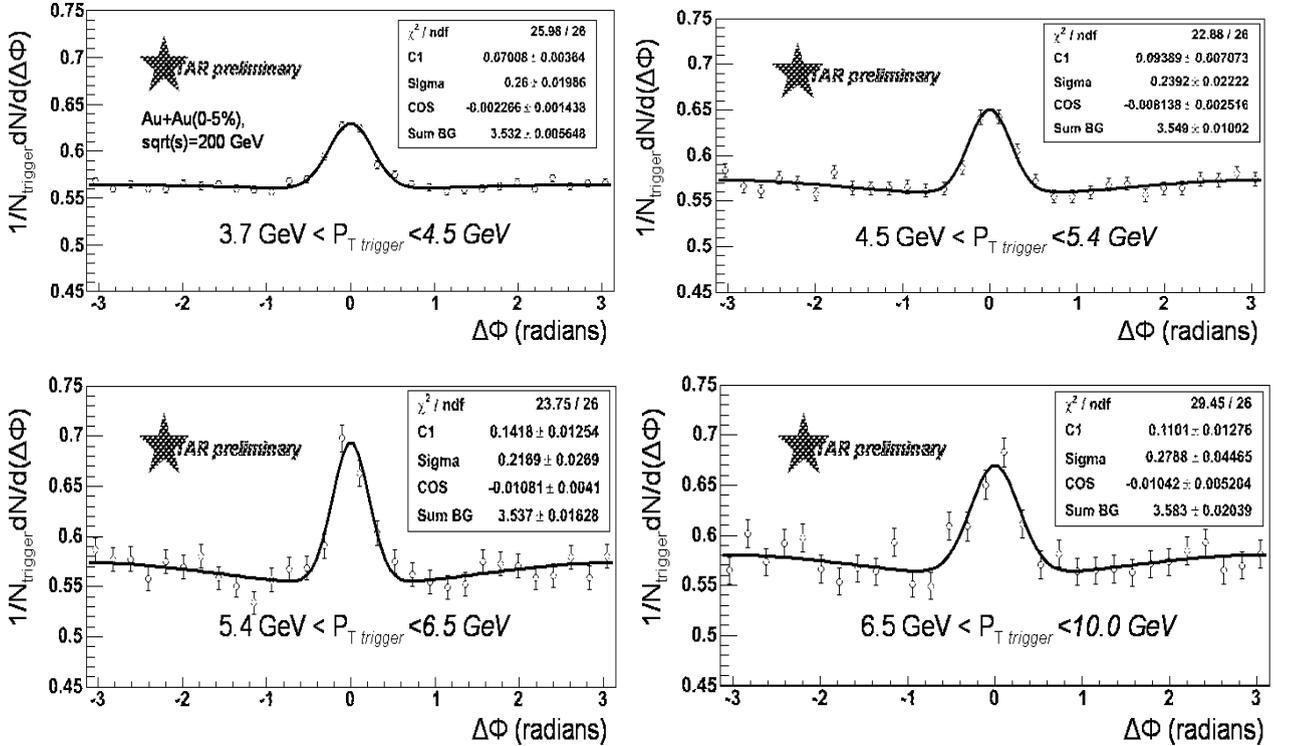

Fig. 4 Di-hadron correlations for charged particles with various ranges, with fits to Eq. (4). In all cases with 2.5 GeV/c <$p_{T,associated}$<3.7 GeV/c,

In Fig. 4 we show correlations of charged hadrons for different $p_T$ cuts. The back side correlation was fitted with a cosine function [5] as shown in equation (3). The fitting parameter ($C_{cos}$) was then compared to the momentum conservation calculation using different particle numbers in the recoil medium as shown in Fig. 5.

$$D(\Delta\varphi) = C_1 e^{-\frac{\Delta\varphi^2}{2\sigma^2}} + \frac{BG}{2\pi} + C_{cos}\cos(\Delta\varphi) \quad (4)$$

As seen in Fig. 4 and 5 both the shape and the magnitude are in agreement with momentum conservation over a large part of the fireball. This suggests that the energy lost of the jets is likely to happen at an early stage of the collision before hadronization.

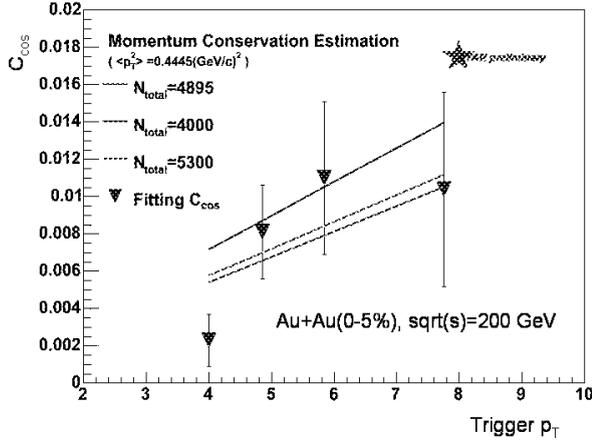

Fig. 5 Fitting parameter compare to estimation from momentum conservation

The redistribution of the quenched energy over such a large portion of the fireball is also consistent with thermalization at the partonic level. The sensitivity of the shape of the measured back side correlation to a non-thermal redistribution of the energy, i.e. momentum conservation over a small number of particles in and near the jet cone, is under investigation by studying the effect with lower $p_T$ limits and for different centralities.

### III. SUMMARY

In this paper, we have shown that the trends of the suppression of the correlation for Lambdas are slightly different from Anti-Lambdas, which could indicate sensitivity in our identified particle jet studies to quenching or production mechanism effects. We have found that the particle azimuthal distribution in the region opposite to the trigger particle is consistent with estimates based on momentum conservation. It indicates that the energy of the quenched jet has been redistributed over many particles in the system, which is consistent with thermalization at the partonic level.